\documentclass[onecolumn]{elsart3p}

\usepackage{graphics}
\usepackage{graphicx}
\usepackage{dcolumn}
\usepackage{amsfonts}
\usepackage{amssymb}

\newcommand{\ar}{\arrowvert}
\newcommand{\ra}{\rangle}
\newcommand{\la}{\langle}

\newcommand{\ov}{\overline}

\newcommand{\be}{\begin{equation}}
\newcommand{\ee}{\end{equation}}
\newcommand{\ba}{\begin{eqnarray}}
\newcommand{\ea}{\end{eqnarray}}

\newcommand{\pa}{\partial}

\begin{document}

\begin{frontmatter}



\title{Bulk viscosity of low-temperature strongly interacting matter}


\author{Antonio Dobado, Felipe J. Llanes-Estrada and Juan M. Torres-Rincon}
\address{Departamento de F\'{\i}sica Te\'orica I, Universidad
Complutense de Madrid, 28040 Madrid, Spain}

\begin{abstract}
We study the bulk viscosity of a pion gas in unitarized Chiral Perturbation Theory at low and moderate temperatures, below any phase transition to a quark-gluon plasma phase. \\
We argue that inelastic processes are irrelevant and exponentially suppressed at low temperatures. Since the system falls out of chemical equilibrium upon expansion,
a pion chemical potential must be introduced, so we extend the existing theory to include it.
We control the zero modes of the collision operator and Landau's conditions of fit when solving the Boltzmann equation with the elastic collision kernel. \\
The dependence of the bulk viscosity with temperature is reminiscent of the findings of Fern\'andez-Fraile  and G\'omez Nicola \cite{FernandezFraile:2008vu}, while the numerical value is closer to
that of Davesne \cite{Davesne:1995ms}. In the zero-temperature limit we correctly recover the vanishing viscosity associated to a non-relativistic monoatomic gas.
\end{abstract}

\end{frontmatter}

\section{Motivation}

Bulk viscosity $\zeta$ is responsible for the equilibration of a system subject to dilatation or compression perturbations. Although it usually is much smaller than the more common shear viscosity $\eta$, it is an interesting probe of the loss of dilatation invariance~\cite{Karsch:2007jc} through the trace anomaly in quantum field theory. In the context of relativistic heavy ion collisions, bulk viscosity can be experimentally accessed for example by means of fluctuations of the two-point correlation of the energy-momentum tensor~\cite{Dobado:2011wc}.
It has been computed in the asymptotically high temperature regime of Quantum Chromodynamics~\cite{Arnold:2006fz}, and found to be governed there by inelastic processes.

However, published investigations concerning bulk-viscosity in the low-energy pion gas (the opposite, low-temperature limit of strongly interacting matter at zero baryon number) yield contradictory results. While the Chiral Perturbation Theory based computation of Fern\'andez-Fraile and G\'omez Nicola features a clear maximum of $\zeta$ associated with the scale-violating pion mass~\cite{FernandezFraile:2008vu}, this is absent from earlier studies~\cite{Davesne:1995ms,Prakash:1993bt} based on empirical phase-shifts, that also differ in the numeric values of $\zeta$.

Neither of these calculations carefully identifies the zero-modes of the collision operator that might change the temperature counting of the viscosity.
A chiral limit calculation has however been reported~\cite{Chen:2007kx} that studies these zero modes in elastic scattering.

Further, none of these works properly emphasizes the role of the Landau-Lifschitz constraint, that is however properly treated in a linear sigma model computation~\cite{Chakraborty:2010fr}.

Given the situation, we feel it is worth to settle the issue of the
bulk viscosity of a low-temperature pion gas by adopting the best
features of  the various existent approaches. In this article we
provide a Unitarized ChPT-based computation at the physical pion
mass $m_\pi$, where the interaction kernel reproduces the physical
phase-shifts, and where the zero modes and Landau-Lifschitz
constraints are correctly implemented.

Since we find that elastic processes dominate the low-temperature bulk viscosity, pion number is conserved in the relevant time-scales, and the need to introduce in the theory a pion chemical potential $\mu$ arises, which is our additional contribution to the discussion. We also want to emphasize that the relativistic chemical potential satisfies $\mu \le m_\pi$
instead of $\mu\le 0$ in the non-relativistic theory. Thus, some of prior work with $\mu=0$ might actually not be easily mapped to the non-relativistic limit (where it would correspond
to an artificial $\mu=-m_{\pi}$ if care was not exercised in subtracting $m_{\pi}$ from the energy), finding a divergence of $\zeta(T)$ at vanishing temperature that might be purely academic.

We make no attempt to extend the validity of the approach to higher temperatures where strangeness is an active degree of freedom and our considerations should be understood as valid up to temperatures in the range 120-140 MeV at most, although we may plot results at slightly larger temperatures for archival and comparison purposes.

\section{Dominance of elastic processes} \label{seccion:elastica}

Departures from equilibrium of the one-particle distribution function $f_p=n_p + \delta f_p$ that are associated with system expansion can be parametrized, in linear order of velocity gradients, as
\be
\delta f_p=-n_p (1+n_p) \beta \  \nabla \cdot \mathbf{V} \ A_p\ ,
\ee
with $n_p$ the Bose-Einstein equilibrium distribution function $n_p = (e^{\beta E}-1)^{-1}$, E the relativistic kinetic energy, and $\beta=1/T$ the inverse temperature.
$A_p$ must be an adimensional function of the particle momentum. Therefore, defining the dimensionless variable $x\equiv E_p/m_\pi =\sqrt{1+(p/m_\pi)^2}$, one can expand it as
\be \label{basisofA}
A_p \simeq \tilde{a}_0 + \tilde{a}_1 E_p + \tilde{a}_2 E_p^2 + \dots
\ee
We can conveniently also consider $A_p$ as a function $A(x)$.
From dissipative hydrodynamics, the bulk viscosity is known to be a simple integral of $A_p$
\be  \label{bulk1}
\zeta =  \frac{g}{T} \int \frac{d^3 p}{(2\pi)^3 E_p} n_p (1+n_p) A_p \ \frac{p^2}{3}
\ ,
\ee
with $g=3$ being the pion isospin degeneracy.
Relaxation of $f_p$ towards equilibrium in a dilute gas is governed by the Boltzmann equation $p_{\mu} \pa^{\mu} f = C[f]$ that is linearized in terms of the separation from equilibrium $\delta f$, yielding a linear system in an appropriate polynomial basis, for example that of Eq.~(\ref{basisofA}), resulting in $C \ar A \rangle = \ar S \rangle$.

The collision operator $C$ has parts associated with elastic $\pi\pi\to \pi\pi$ and inelastic $\pi\pi\to \pi\pi\pi\pi$ processes, $C=C_2+C_4+\dots$

Writing explicitly the elastic part $C_2$ and assuming for the time being that the system does not have a well defined particle number (no chemical potential), the linearized Boltzmann's equation with a compressive disturbance has the well-known form
\ba \nonumber \label{boltzmann1}
\ar S\rangle \equiv  \left( \frac{1}{3} p^2 -E_p^2 v^2 \right)  =
\\
 \frac{gE_p}{2n_p(1+n_p)} \int d \Gamma_{12,3p}  (1+n_1)(1+n_2)n_3 n_p [ A_p + A_{p_3} -A_{p_1} - A_{p_2}]\ \ \  +\ \ \  C_4 \ar A\rangle
\ea
with $d\Gamma_{12,3p}$ carrying the elastic interaction and defined below in Eq.~(\ref{intmeasure}), and $v^2$ being the squared sound velocity $\partial P/\partial \rho$.

The zero modes of the elastic collision operator $C_2$ can be identified from the structure $A_p + A_{p_3} -A_{p_1} - A_{p_2}$ and correspond to $A_0=1$ (associated to particle-number conservation since the number of terms in this equation is two positive and two negative due to the $2\to 2$ scattering) and $A_0=x=E_p/m_\pi$ associated to energy conservation (this mode is also present in the inelastic parts of the linearized collision operator).
Thus the collision operator must be inverted in a linear space perpendicular to these modes. Assuming this has been achieved, the linearized Boltzmann equation returns a solution $A_{p\perp}$. However, this solution is not unique as $A_\perp +\tilde{a}_1 x$
will also solve it for arbitrary $\tilde{a}_1$ including both collision operators $C_2+C_4$, and $A_\perp +\tilde{a}_0$ for arbitrary $\tilde{a}_0$ including only $C_2$. These two constants can be specified by the two Landau-Lifschitz constraints (``conditions of fit'') necessary to fully define the reference frame and the particle density, that are
\be
\tau^{00}= \int \frac{d^3 p}{(2\pi)^3} \delta f_p E_p= 0 \ ;
\ \ \ \ \ \   \ \ \ \ \ \ \ \ \
\nu^0=  \int \frac{d^3 p}{(2\pi)^3} \delta f_p =0 \ .
\ee
Substituting $\delta f_p$ in terms of $A$, these can also be written as
\be \label{LandauLifschitz}
 \int \frac{d^3 p}{(2\pi)^3 E_p} n_p (1+n_p) A_p E^2_p =0\ ;
\ \ \ \ \ \   \ \ \ \ \ \ \ \ \
\int \frac{d^3 p}{(2\pi)^3 E_p} n_p (1+n_p) A_p E_p =0
\ee

The Boltzmann equation can then be symbolically written as
\be
(C_2+C_4) \ar A_\perp + A_0 \ra = \ar S \ra \ \ \ \  {\rm{or}}\ \ \ \
C_2 \ar A_\perp \ra + C_4 (\ar A_\perp\ra + \ar A_0\ra) =\ar S\ra
\ee
in view that $A_0$ contains the zero modes of $C_2$.

Now we seek to see which of these terms dominates the inversion of the collision operator at low temperatures. It is clear that, for massive particles, one can use non-relativistic power counting for low-enough temperature, so that $E\simeq m_\pi + p^2/2m_\pi$, with $m_\pi \propto O(1)$, $(E_p-m_\pi) \propto O(k_BT)$, $p\propto \sqrt{m_\pi T}$. Then it is possible that $A_\perp$ be suppressed by a temperature power respect to $A_0$ given that the zero modes involve the lowest powers of the energy $1$, $x$. This is however overcompensated by the exponential suppression of inelastic processes
$C_4/C_2\propto e^{-2m_\pi /T}$ from the additional $n_p$ factors, and an additional two powers of $T$ from chiral suppression of the inelastic amplitude. Therefore, at low temperatures $C_4 \ar A_0\ra $ is negligible respect to $C_2 \ar A_\perp \ra$. In the following we neglect the contribution of $C_4$, that however should become more and more relevant as one approaches the phase transition (or alternatively, when typical CM collision energies are of the order of the inelastic resonance $f_2(1270)$). However the strangeness degree of freedom is active before that energy, so including $C_4$ without attending to the kaons in the bath is not a sensible approach.

We therefore concentrate on elastic processes and proceed to analyze $C_2$ in detail. Since elastic collisions do not change pion number, there is an approximate conservation law that requires the introduction of a chemical potential.

\section{Bulk viscosity in the presence of a pion chemical potential}

In the presence of a pion chemical potential, the left hand side of
the Boltzmann equation~(\ref{boltzmann1}) needs to be modified \be
\left. p_{\mu} \pa^{\mu} f_0 (x) \right|_{\zeta} =\beta n_p (1+n_p)
\left( \frac{1}{3} p^2 -E_p^2 v_n^2  -E_p \kappa^{-1}_{\rho} \right)
\nabla \cdot \mathbf{V},\ee with the inverse compressibility at
fixed density and the sound velocity at constant particle density
defined as \be \kappa^{-1}_{\rho} = \left( \frac{\pa P}{\pa n}
\right)_{\rho}; \quad v_n^2 = \left( \frac{\pa P}{\pa \rho}
\right)_n\ , \ee where $n=n(T,\mu)$ is the particle number density
and $\rho=\rho(T,\mu)$ is the energy density. The source function in
the left hand side of Eq.~(\ref{boltzmann1}) is therefore to be
substituted by $\left( \frac{1}{3} p^2 -E_p^2 v_n^2  -E_p
\kappa^{-1}_{\rho} \right)$.

It is then convenient to employ the Landau-Lifschitz conditions~(\ref{LandauLifschitz})
to add two (vanishing) terms to Eq.~(\ref{bulk1}) to make
the bulk viscosity take an elegant form
\be  \label{bulk2}
\zeta =  \frac{g}{T} \int \frac{d^3 p}{(2\pi)^3 E_p} n_p (1+n_p) A_p \ \left( \frac{p^2}{3} - E_p^2 v_n^2 - E_p \kappa^{-1}_{\rho} \right) \ .
\ee
Since the collision operator is a linear operator acting on $A$, it is also convenient to define a scalar product in the function space given by the measure
\be
d\mu= d^3 p \frac{1}{4\pi m_\pi^2} \frac{1}{E_p} n_p (1+n_p),
\ee
and a basis of the function space. After much trial and error we have found practical
to employ a basis that is not totally orthogonal, but where the first and third polynomials of $x=E/m_\pi$ are given by
\be
P_0(x)\equiv 1\ ;  \ \ \ \ \ \
P_2(x)\equiv \left( \frac{1}{3} - v_n^2 \right) x^2 - \frac{\kappa^{-1}_{\rho}}{m_\pi} x- \frac{1}{3}\ .
\ee
All other polynomials $P_1(x)$, $P_3(x)$ and higher  are then chosen by standard orthogonalization to these two, and monic, that is, in $P_n$ the coefficient of $x^n$ (maximum power of $x$) is 1. The advantages of these basis are that the first two polynomials $P_0$ and $P_1$ span the zero modes of the collision operator, and the second, $P_2$, is the source $\ar S\ra$ function, bringing about certain simplification.
(We should like to stress again that for $i\ge 1$ the $P_i$ form an orthogonal, monic basis).

Equation (\ref{bulk2}) for the bulk viscosity is then the projection of the solution to Boltzmann's equation over the source,
\be
\zeta = \frac{gm_\pi^4}{2 \pi^2 T} \langle A(x) | P_2(x) \rangle\ .
\ee

\section{Solution of Boltzmann equation and satisfaction of constraints}

We seek the coefficients of the expansion
\be
A(x) = \sum_{n=0}^{\infty} a_n P_n (x)
\ee
that satisfy both the quantum Boltzmann equation $C_2 \ar A\ra =\ar S \ra$ and the Landau-Lifschitz constraints $\la E_p \ar A \ra=0$, $\la E_p^2 \ar A \ra=0$

By truncating the polynomial basis $\left\{ P_i (x) \right\}$ to a finite number of polynomials, one can represent the linearized Boltzmann equation as a symmetric
linear system that can be inverted on a computer, in terms of $C_{ij}=\la P_i \ar C \ar P_j \ra=C_{ji}$ that is the matrix element of the collision operator in this basis.

\be
\left(
\begin{tabular}{c|ccc}
0     & 0 & 0 & \dots \\
\hline
0     & $C_{22}$ & $C_{23}$ & \dots \\
0     & $C_{23}$ & $C_{33}$ & \dots \\
\dots & & \dots &
\end{tabular}
\right)\left(
\begin{tabular}{c}
${\mathfrak{a}_1}$ \\
${\mathfrak{a}_2}$ \\
${\mathfrak{a}_3}$ \\
\dots
\end{tabular}
\right) =
\left(
\begin{tabular}{c}
0 \\
$S_2$ \\
0 \\
\dots
\end{tabular}
\right) \ .
\ee
This system is compatible but indeterminate. Inclusion of $P_0(x)=1$ in the space where $C$ is to be inverted makes the system incompatible, since it is a zero mode and the source
has non-zero projection $\la P_0 (x) | S \ra \neq 0$. It is not possible to find a basis in which $P_0(x)$ and $P_1(x)$ (spanned by the two zero modes) are both orthogonal to $P_2(x)=\ar S\ra $.
In the function space ${P_1\dots P_n}$ that we keep, the source, by construction, has only a projection over $P_2$, $S_2=\la S\ar P_2\ra$. \\
As for the indeterminacy, note that $\mathfrak{a}_1$ is so far completely arbitrary.

 The first step is thus to invert $C$ in a subspace $\left\{ P_2\dots P_n \right\}$. If one truncates at $n=2$, for example, then $\mathfrak{a}_2= C_{22}^{-1} S_2$.

We then identify $a_n=\mathfrak{a}_n$, for $n\ge 2$, and we fix $a_0$ and $a_1$ by demanding satisfaction of Eq.~(\ref{LandauLifschitz}). For example, truncating at $n=2$,
we would find
\ba a_0 \langle x^2 | P_0 \rangle + a_1 \langle x^2 | P_1 \rangle &=& -a_2 \langle x^2 | P_2 \rangle \\ \nonumber
a_0 \langle x | P_0 \rangle +\  a_1 \langle x | P_1 \rangle &=& -a_2 \langle x | P_2 \rangle
\ea
For higher $n$ it is a matter of solving a more complex $2\times 2$ linear system, task that we perform easily with a computer program. Essentially all the computer time is spent in obtaining the $C_{ij}$ matrix elements by means of Montecarlo integration over the multidimensional space of particle momenta, with the linear algebra in this section costing negligibly small CPU effort.

Note that the space where we actually solve the Boltzmann equation is not perpendicular to the zero modes (our discussion in section~\ref{seccion:elastica} above was too naive). However, the inversion is well conditioned and convergence in $n$ very fast because there is a non-zero projection of both zero modes that is perfectly perpendicular to the inversion function space, so that no eigenvalue ever vanishes.

The resulting $A(x)$ can then be substituted in either the Boltzmann equation or
the Landau-Lifschitz conditions and satisfies both uniquely.

\section{Low-energy pion scattering}

The interaction between pions ($1+2 \rightarrow 3 +p$) appears in the squared matrix element $\ar T\ar^2$
\be \label{intmeasure}
d\Gamma_{12,3p}
= \frac{1}{2E_p } \ov{\ar T\ar^2}\prod_{i=1}^3 \frac{d{\bf k}_i}{(2\pi)^32E_i}
(2\pi)^4 \delta(k_1+k_2-k_3-p)
\ee
averaged over isospin and partial waves.
We employ the Inverse Amplitude Method (IAM)~\cite{Dobado:1989qm} to represent $\ar T\ar^2$. The interaction is exactly the same as in~\cite{Dobado:2003wr} and well documented in the literature, in particular we employ the fits from Ref.~\cite{GomezNicola:2001as}.
The key of the method is to employ a dispersive analysis to describe pion-pion scattering in which the function in the dispersive integral is not the partial wave itself but its inverse. The subtraction constants and left cut are approximated in Chiral Perturbation Theory, while the right cut is known exactly in the elastic region (and to a very good approximation up to about 1.2 GeV). The resulting IAM formula is then written down in terms of the Chiral Perturbation Theory expansion for the partial waves, $t\simeq t_{2}+t_{4}+\dots $, to name
\be
t(s) \simeq \frac{t_2^2(s)}{t_2(s)-t_4(s)} \ .
\ee
Those partial waves with $(IJ)=(00)$, $(11)$, $(20)$ are used to construct $\ar T\ar^2$. Agreement with experimental data on pion scattering is very good and the theoretical constraints of unitarity and chiral symmetry are properly implemented.

The low-energy pion resonances in the scalar ($\sigma$) and vector ($\rho$) channels naturally appear in this description, although not as elementary degrees of freedom, but as dynamically generated resonances.

\section{Results and Discussion}

\begin{center}
\begin{figure}[th]
\parbox{8cm}{\includegraphics[width=8cm,angle=0]{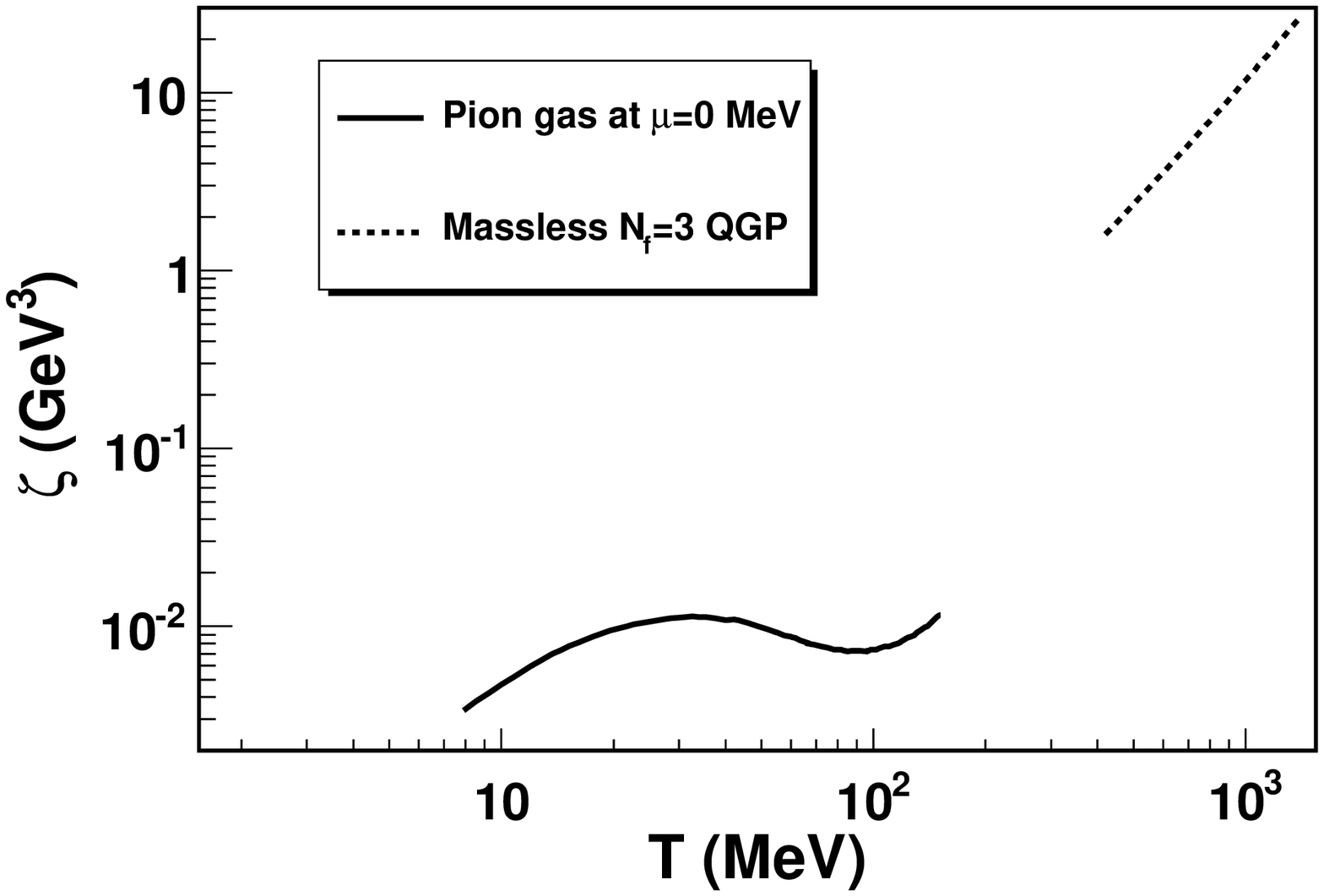}}
\parbox{8cm}{\includegraphics[width=8cm,angle=0]{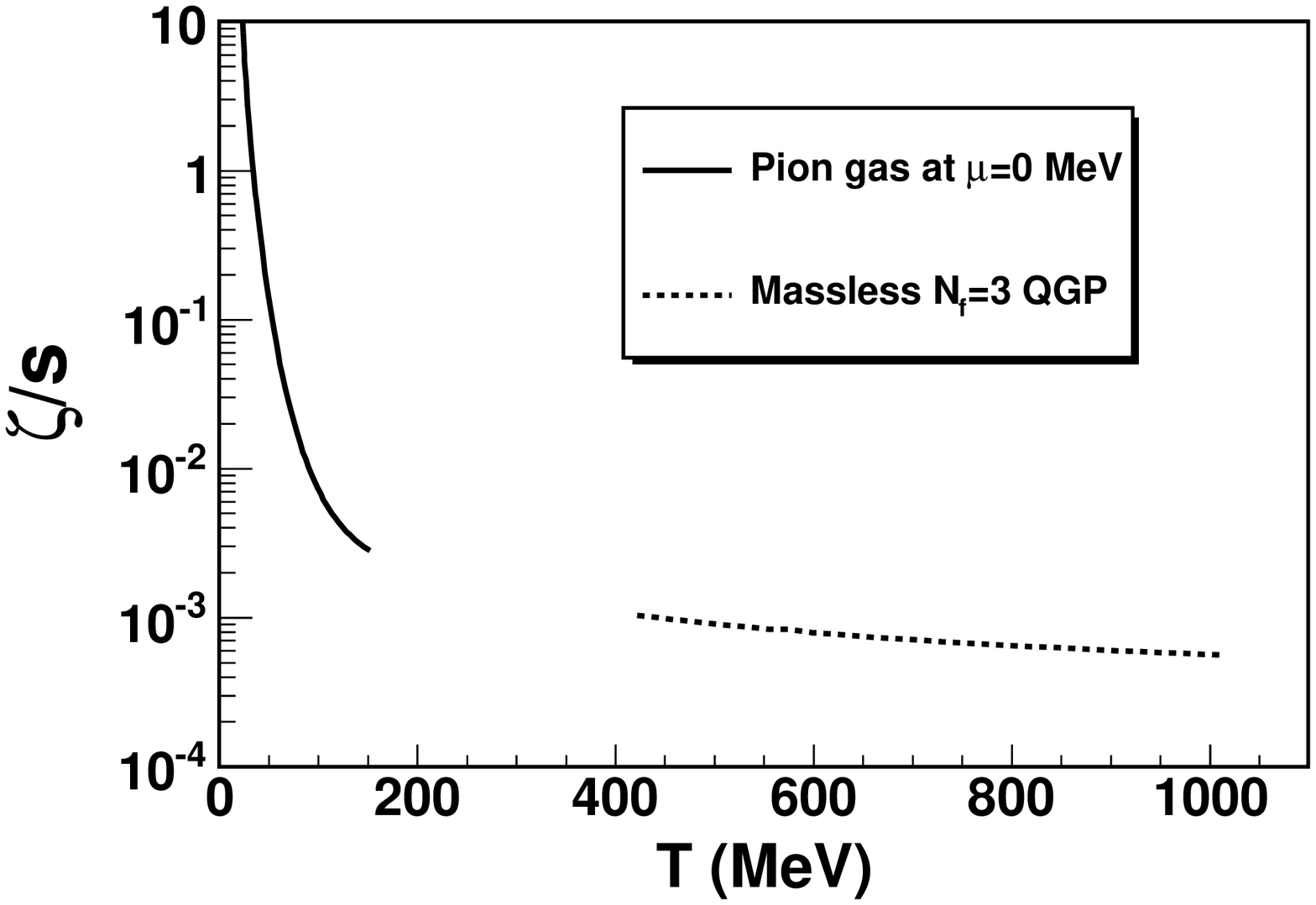}}
\caption{Bulk viscosity of a pion gas in the Inverse Amplitude Method (left) and ratio of bulk viscosity to entropy density (right). In addition we plot the asymptotic quark-gluon plasma estimate of~\cite{Arnold:2006fz}.
\label{mainresult1}}
\end{figure}
\end{center}

Our numerical computation for the bulk viscosity of the pion gas, based on elastic processes, is plotted in Figure~\ref{mainresult1} (left panel). We observe a broad maximum at low temperatures, consistent with $\zeta=0$ at zero temperature (a non-relativistic prediction for the monoatomic gas), and a clear growth of $\zeta$ at larger temperatures approaching the phase transition to the quark-gluon plasma. Starting at a higher, 400 MeV temperature, we have plotted also the asymptotic estimate for $\zeta$ in that phase~\cite{Arnold:2006fz}. In the right panel we also give these results quotiented by the entropy density $s$, since this is a dimensionless number that characterizes the sound wave attenuation.
For SU(3) and $N_f=3$ we have~\cite{Csernai:2006zz}, in the high temperature regime

\be \frac{\zeta}{s} = 2 \cdot 10^{-4} \frac{g^4}{\log \left(\frac{6.344}{g} \right)} \ , \qquad g^{-2} (T)=\frac{9}{8\pi^2} \log \left( \frac{T}{30 \textrm{ MeV} } \right) + \frac{4}{9\pi^2} \log \left( 2 \log \left( \frac{T}{30 \textrm{ MeV}} \right) \right) \ . \ee
This ratio we find ourselves in the pion gas to be monotonously decreasing as a function of temperature.

\begin{center}
\begin{figure}[th]
\parbox{8cm}{\includegraphics[width=8cm,angle=0]{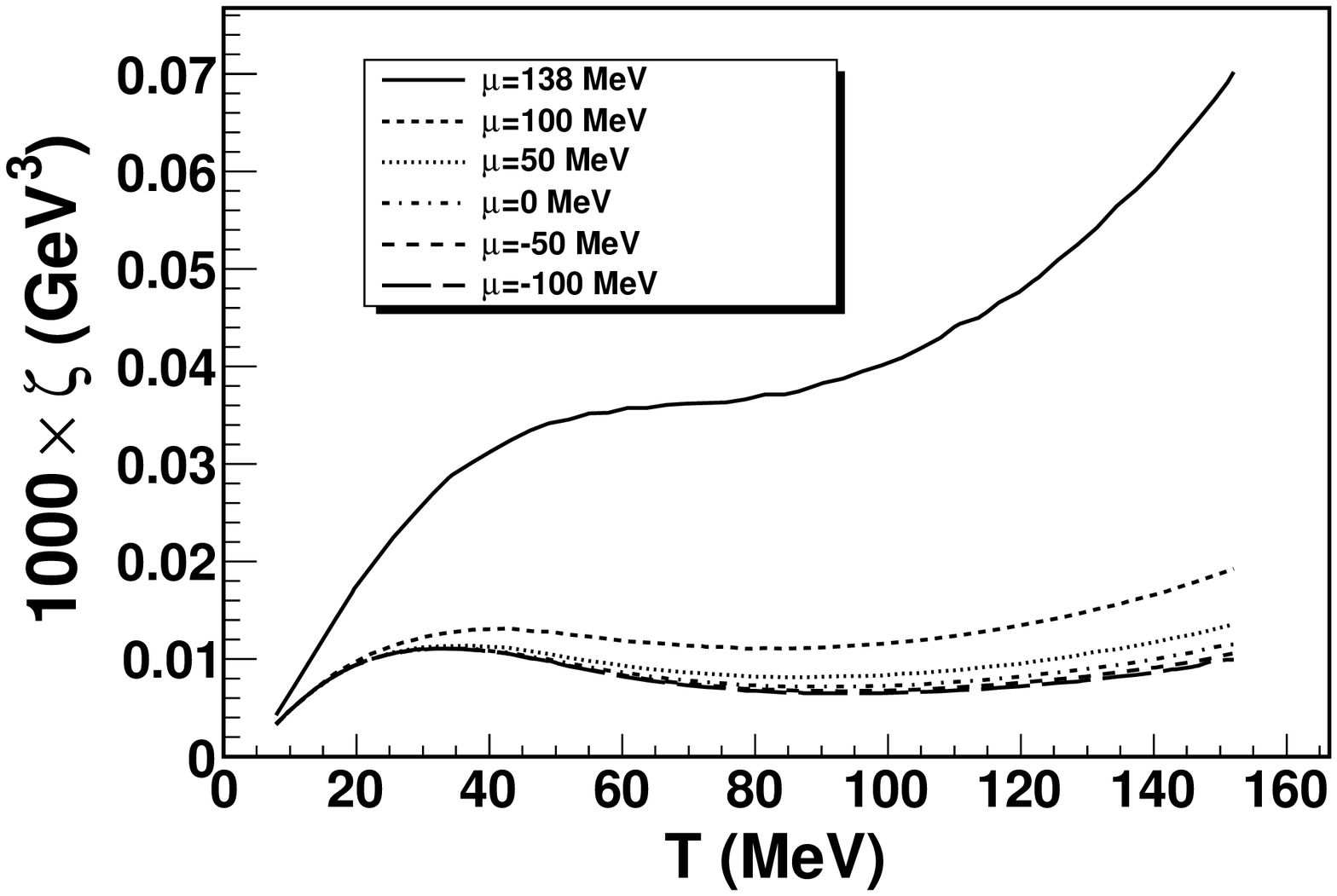}}
\parbox{8cm}{\includegraphics[width=8cm,angle=0]{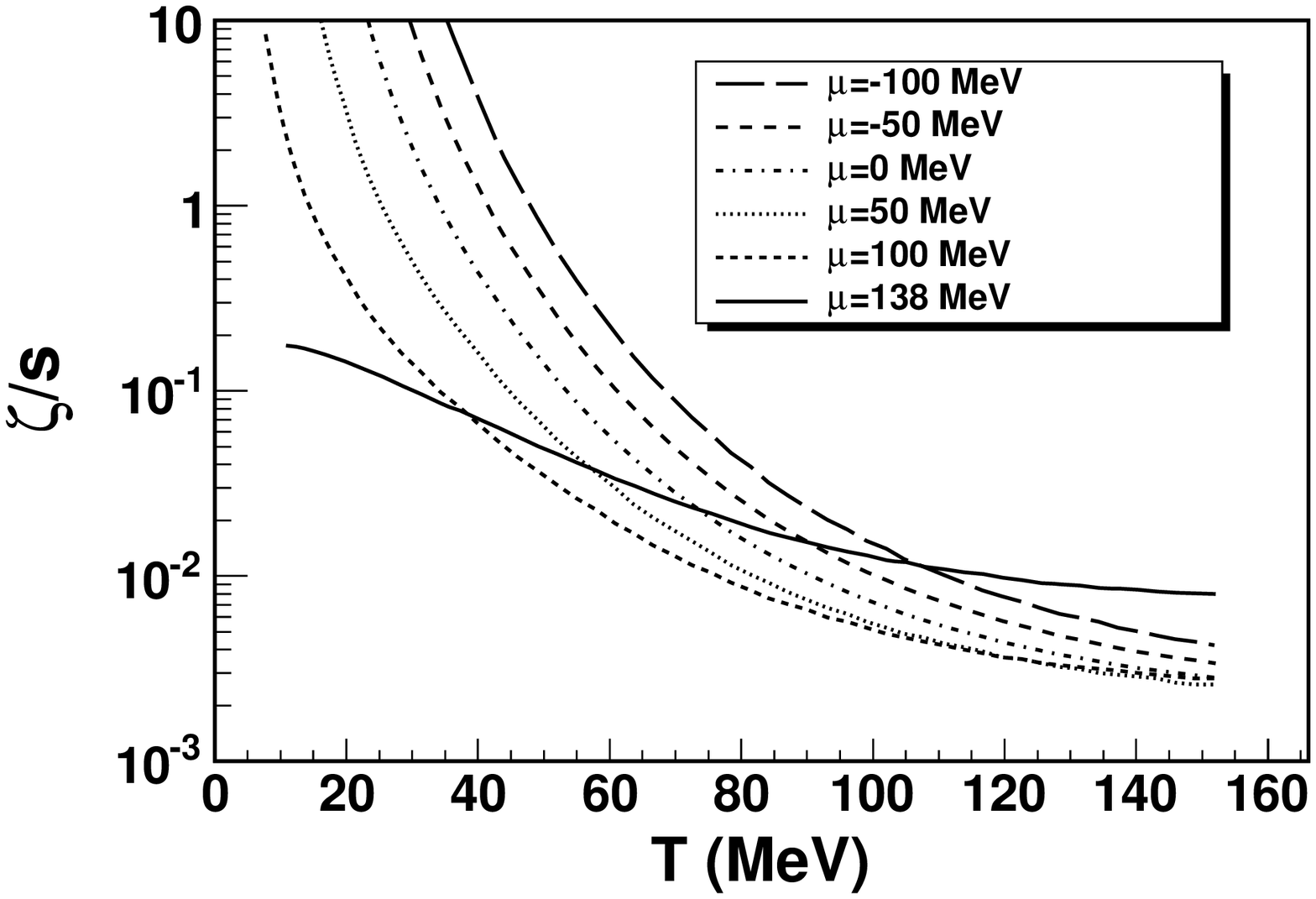}}
\caption{Same as in Figure~\ref{mainresult1} as a function of the chemical potential for the pion gas.
\label{mainresult}}
\end{figure}
\end{center}

Since we have contributed for the first time an evaluation that allows for the chemical potential, we also repeat the plot in Figure~\ref{mainresult1} for various values of $\mu$ in Figure~\ref{mainresult} (for the pion gas only).
\\
We should like to remark again that the natural end-value is
$\mu=m_\pi$ which corresponds to chemical equilibrium. Increasing
$\mu$ generally yields a larger viscosity.

\begin{center}
\begin{figure}[th]
\parbox{8cm}{\includegraphics[width=8cm,angle=0]{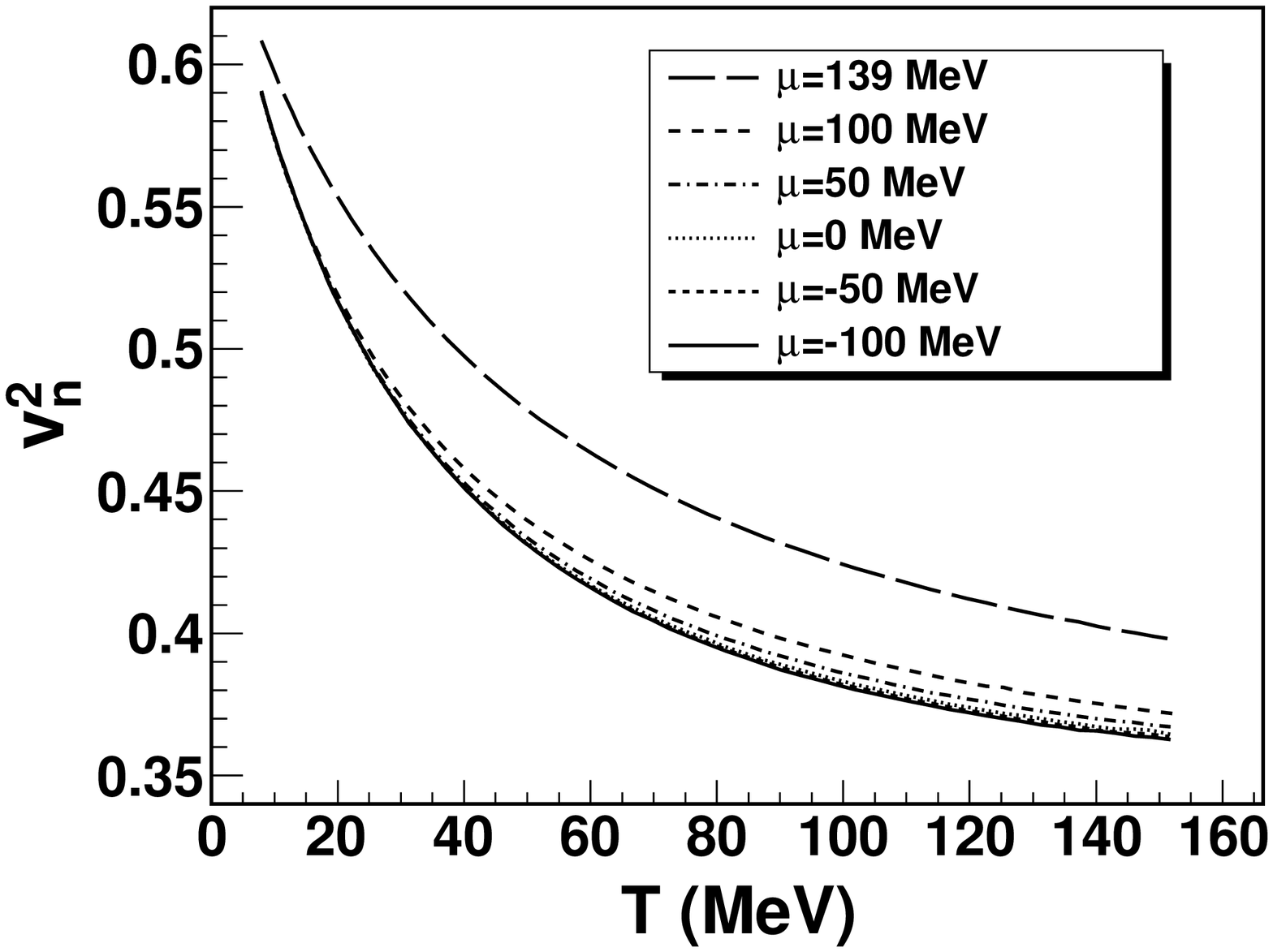}}
\parbox{8cm}{\includegraphics[width=8cm,angle=0]{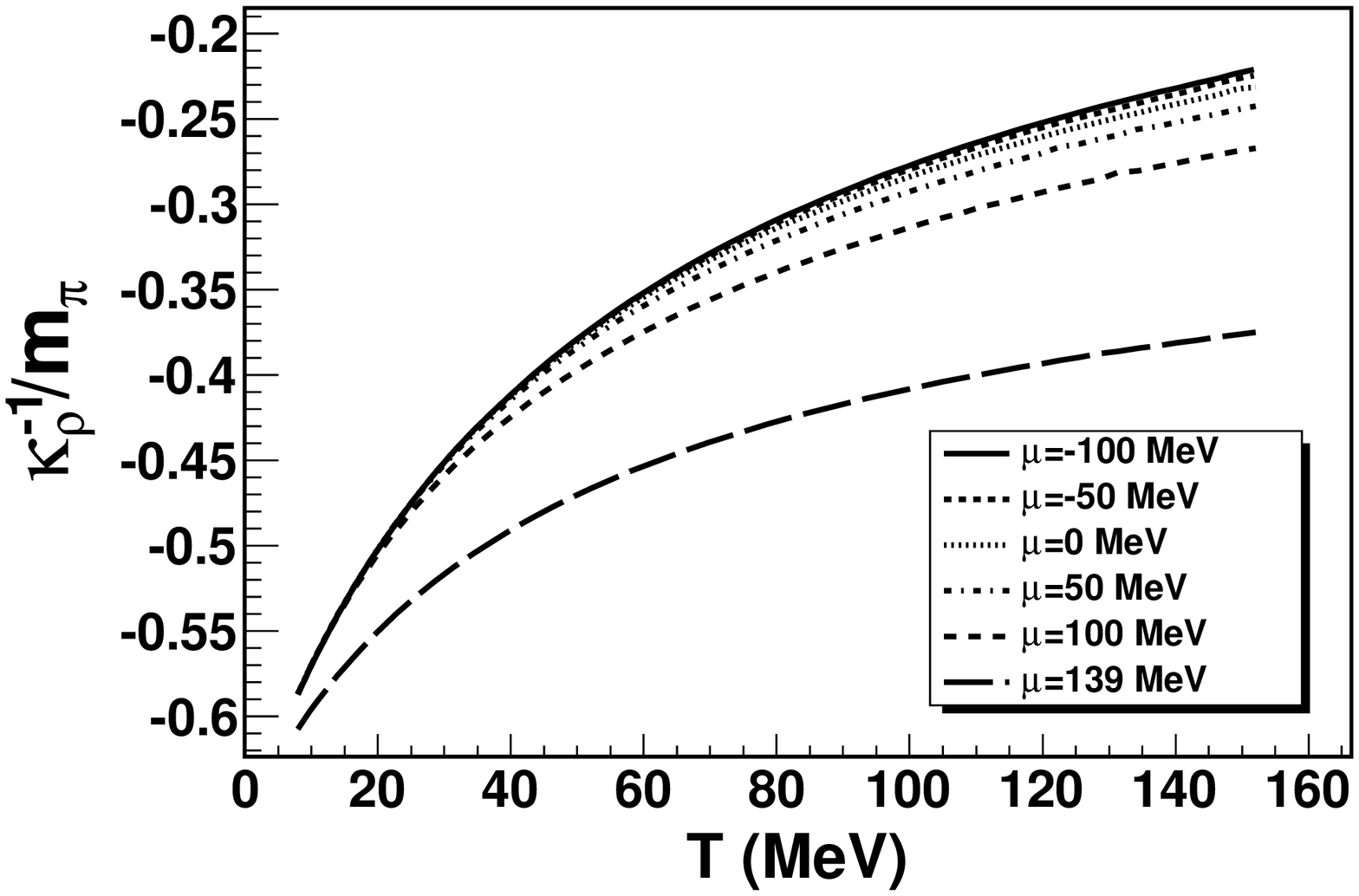}}
\caption{Sound speed $v_n$ at fixed particle density  (left figure) instead of fixed entropy as conventionally used when $\mu=0$. Inverse compressibility (right figure) $\kappa^{-1}_{\rho}$ in units of the pion mass.
\label{termofig}}
\end{figure}
\end{center}

Since this calculation employs two quantities that are not widely used in the literature
(the sound speed at fixed particle density as appropriate for the fixed-chemical potential calculation, and the inverse compressibility), we plot them also in Figure~\ref{termofig} for the reader's convenience.

\begin{center}
\begin{figure}[t]
\centerline{\parbox{10cm}{\includegraphics[width=10cm,angle=0]{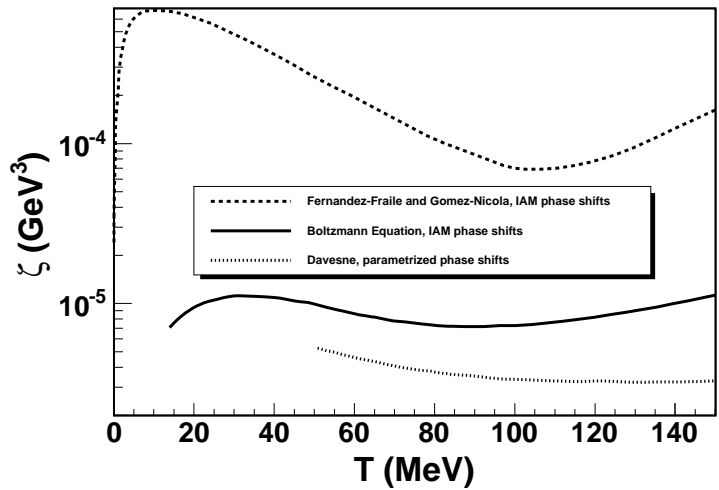}
\vspace{0.01cm}}}
\caption{Comparison of our computation with prior evaluations at $\mu=0$. The solid line is our computation with the Inverse Amplitude Method at zero chemical potential. The numerical result is similar to Davesne's evaluation based on a phenomenological parametrization of the physical pion phase-shifts~\cite{Davesne:1995ms} (dotted line at the bottom).
Extending the computation to lower temperatures, we find a maximum and then a return to small values of the bulk viscosity, in agreement with Fern\'andez-Fraile and G\'omez Nicola in chiral perturbation theory (that is however numerically a little too large) , and the non-relativistic vanishing behavior of $\zeta$ for a monoatomic gas.
 \label{comparison}}
\end{figure}
\end{center}

\begin{center}
\begin{figure}[t]
\centerline{\parbox{10cm}{\includegraphics[width=10cm,angle=0]{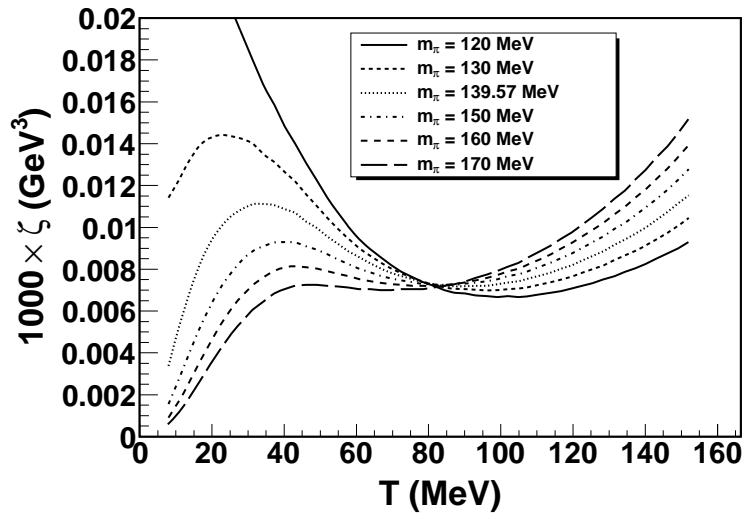}
\vspace{0.01cm}}}
\caption{Pion mass dependence of the bulk viscosity. The  peak at moderate temperature is seen to be very sensitive to the pion mass.
 \label{massdependence}}
\end{figure}
\end{center}

We also wish to compare our numerical results with prior approaches based on the elastic pion-pion interactions. We have chosen two works. The first~\cite{Davesne:1995ms} employs a pion scattering amplitude that fits the experimental phase-shifts but has no connection to chiral perturbation theory. The author does not plot the viscosity for low temperatures. Where he provides his data, our calculation is numerically similar but somewhat higher. \\
The second computation~\cite{FernandezFraile:2008vu} is a field theory evaluation based on a certain ladder resummation, and is numerically off our result based on the physical phase shifts. However, the qualitative features and saliently the low-temperature limit coincide with our findings.\\
Those authors also mention that the first peak is due to the scale provided by the pion mass, $m_\pi$. Therefore we vary this mass in our computer program, and plot the result in Figure~\ref{massdependence}. While true that the position of the maximum reacts to the change in the pion mass, it does so in an unpredicted way. Upon increase of the pion mass, the bulk viscosity actually decreases, feature not trivially in accordance to the expectation of proportionality to the scale-invariance anomaly $\zeta\propto \Theta^\mu_\mu$. The position of the peak does shift to higher temperatures (qualitatively in agreement with the scale-invariance violations appearing at $T\simeq m_\pi$). \\
Upon increasing the chemical potential this peak is seen to reduce and even disappear at $\mu=m_\pi$. We temptatively conclude that this peak might be caused by a mismatch and interplay of the two parameters $\mu$ and $m_\pi$.

To summarize, we have provided a wrap-up of what is known from the bulk viscosity in the low-energy pion gas, and find the transport coefficient to be dominated by elastic processes. \\
With an independent Boltzmann calculation we have extended the computation to include finite chemical potential. We find qualitative agreement with prior estimates based on those elastic processes. We look forward to heavy-ion experiments being able to access this interesting transport coefficient proportional to the trace anomaly in the underlying quantum field theory. In a separate publication~\cite{Dobado:2011wc} we are proposing an observable that can be used experimentally.

Note added: In the final phases of composition of this work we have been made aware of a computation~\cite{Lu:2011df} of the contribution of inelastic pion scattering processes to the scattering rate for a pion in the medium. This contribution to the rate, that does not suffer from the particle-conservation zero mode, is found to be exponentially suppressed in agreement with our argument.

\emph{Work supported by grants FPA 2008-00592, FIS2008-01323, FPA2007-29115-E, PR34-1856-BSCH, UCM-BSCH GR58/08 910309, PR34/07-15875 (Spain).
JMT-R is a recipient of an FPU scholarship.}


\end{document}